\documentclass[11pt,twoside]{article}
\usepackage{CAGN2019}
\usepackage{graphicx}

\usepackage[T1]{fontenc} 

\usepackage{latexsym}
\usepackage{verbatim}

\usepackage{ifpdf}  
\ifpdf  
      \DeclareGraphicsExtensions{.pdf,.png,.jpg}  
\else  
      \DeclareGraphicsExtensions{.eps}  
\fi 

\begin{document}

\vskip 1.0cm
\markboth{O.~Cavichia and M.~Moll\'a.}{The formation of the Galactic bulge}
\pagestyle{myheadings}
%
%
\vspace*{0.5cm}
\parindent 0pt{Poster}


\vspace*{0.5cm}
\title{The formation of the Galactic bulge in an inside-out scenario}

\author{O.~Cavichia$^1$ and M.~Moll\'a$^2$}
\affil{$^1$Instituto de F\'isica e Qu\'imica, Universidade Federal de Itajub\'a, Av. BPS, 1303, 37500-903, Itajub\'a-MG, Brazil\\
$^2$Departamento de Investigaci\'on B\'asica, CIEMAT, Avda. Complutense 40, E-28040 Madrid, Spain}

\begin{abstract}
Chemical evolution models (CEM) are important tools to understand the formation and evolution of the components of the Milky Way Galaxy and other galaxies in the universe. The Galactic bulge is the only galaxy bulge that can be resolved and can be studied with exquisite details. In this way, the bulge metallicity distribution function (MDF) can be traced for different regions within the bulge and can give us clues about the bulge formation scenario. In this work we have assumed an inside-out formation for the Galactic bulge and using a CEM we were able to compute the chemical evolution in nine different radial regions, from 0 to 2 kpc, in steps of 0.25 kpc . The preliminary results show that in the inner regions of the bulge the MDF is skewed to higher metallicities, while at the outer regions there is a metal rich component but also a metal poor component much more extended than in the inner regions. These results may explain the metallicity gradient observed in the Galactic bulge.


\end{abstract}

\section{Introduction}

Chemical evolution models (CEM) have been extensively used to study the formation and evolution of the Milky Way Galaxy (MWG) and other galaxies in the universe. The outputs of a CEM depend on the physics adopted, e.g. the stellar yields, initial mass function, star formation rate and gas infall law. One of the most important constraints for CEM is the metallicity distribution function (MDF). In the results of the first studies, the Galactic bulge (GB) appeared as composed mostly by old stars and the MDF was well characterized by a broad single population. Nowadays this picture has changed and the observations \citep[e.g.][]{rojas-arrigada14} have shown that the GB MDF is much more complex than previous thought, composed by at least two populations.  \citet{zoccali17}, hereafter Z17, have demonstrated that the two components present a different spatial distribution, with the metal poor population concentrated near the Galactic centre (GC). 

In this regarding, the CEM can give important clues to understand the formation and evolution of the components of the MWG and other galaxies in the universe. Since the MDF in a CEM is sensible to the time-scale, in this work we use the GB MDF to constraint the time-scale of the GB formation in order to test a scenario where the it is formed inside-out.

\section{Method}
\label{method}

In the case of spiral galaxies, the central bulge brightness profile can be well described by a S\'{e}rsic function \citep{sersic68}. By assuming that the mass distribution follows the light distribution, the GB total mass can be calculated by integrating this profile. Using the properties of the gamma function, we may show that the GB mass enclosed in a radius $R_k$ of the $k$-th radial region can be calculated as: 
\begin{equation}
M_B(<R_k)=M_B \left(1-e^{-\xi}\sum_{i=0}^{2n-1}\frac{\xi^i}{i!}\right), 
\end{equation}
where $M_B=\pi I_0h^2(2n)!$ is the GB total mass and $\xi = (R_k/h)^{1/n}$. In these equations, $I_0$ is the central intensity, $h$ is a scale radius and $n$ the S\'{e}rsic index. Values used for these parameters are: $n=4$, $k=7.66925$, $h=1.67944\times 10^{-4}$ and $M_B=1.7891\times 10^8 {\rm M}_\odot$. We have assumed an inside-out formation for the GB and, using a CEM described in \citet{molla15, molla16}, we were able to compute the chemical evolution in nine different radial regions, from 0 to 2 kpc, in steps of 0.25 kpc . The time-scale to form each radial region of the GB is calculated with the initial mass available in the halo to form the region and the final mass in the respective $k$-th region: $\Delta M_{B_k}=M_B(< R_{k})-M_B(<R_{k-1})$ \citep[see][for details]{molla15}.

\begin{figure}[!ht]  
\begin{center}
\includegraphics[height=4.5cm]{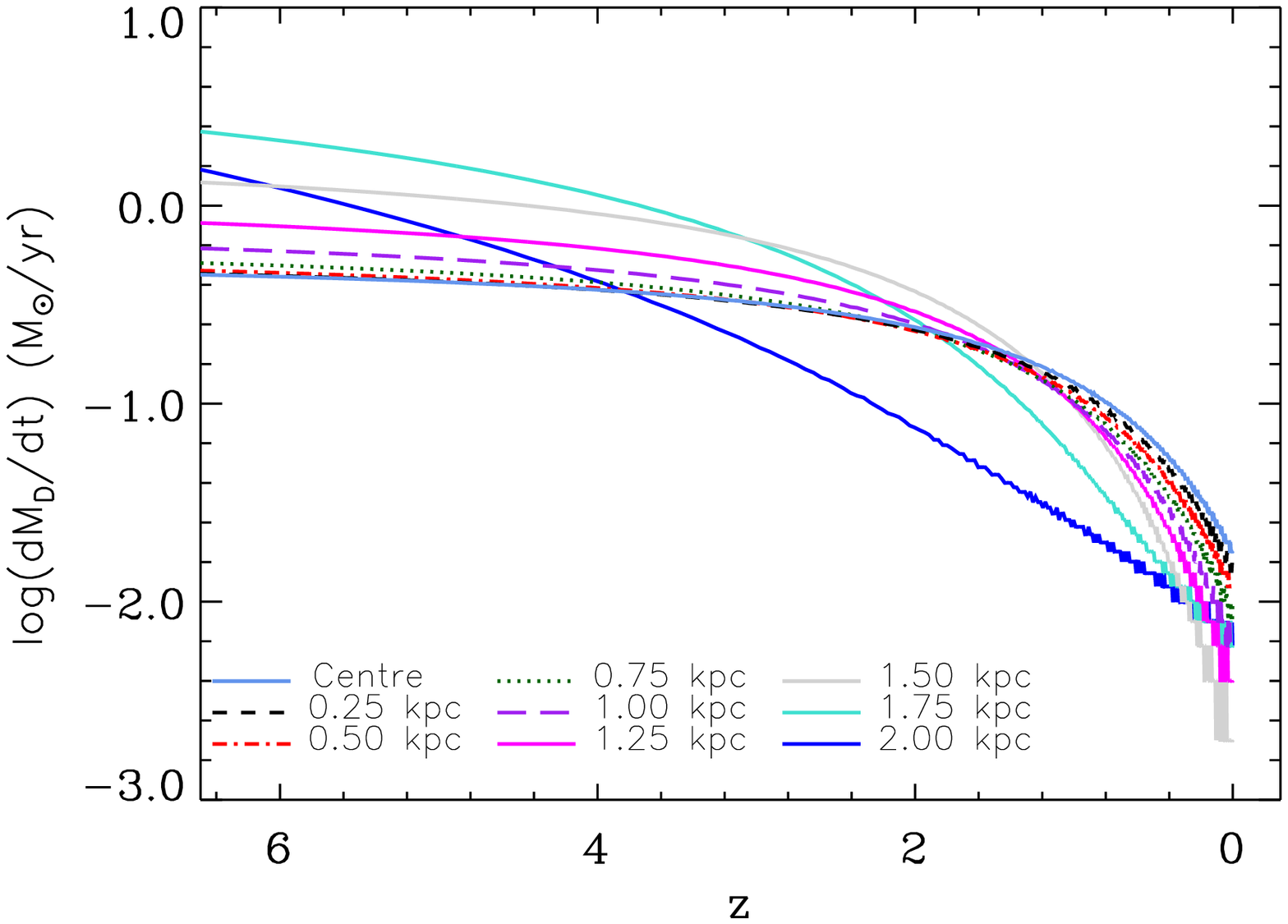}
\includegraphics[height=4.5cm]{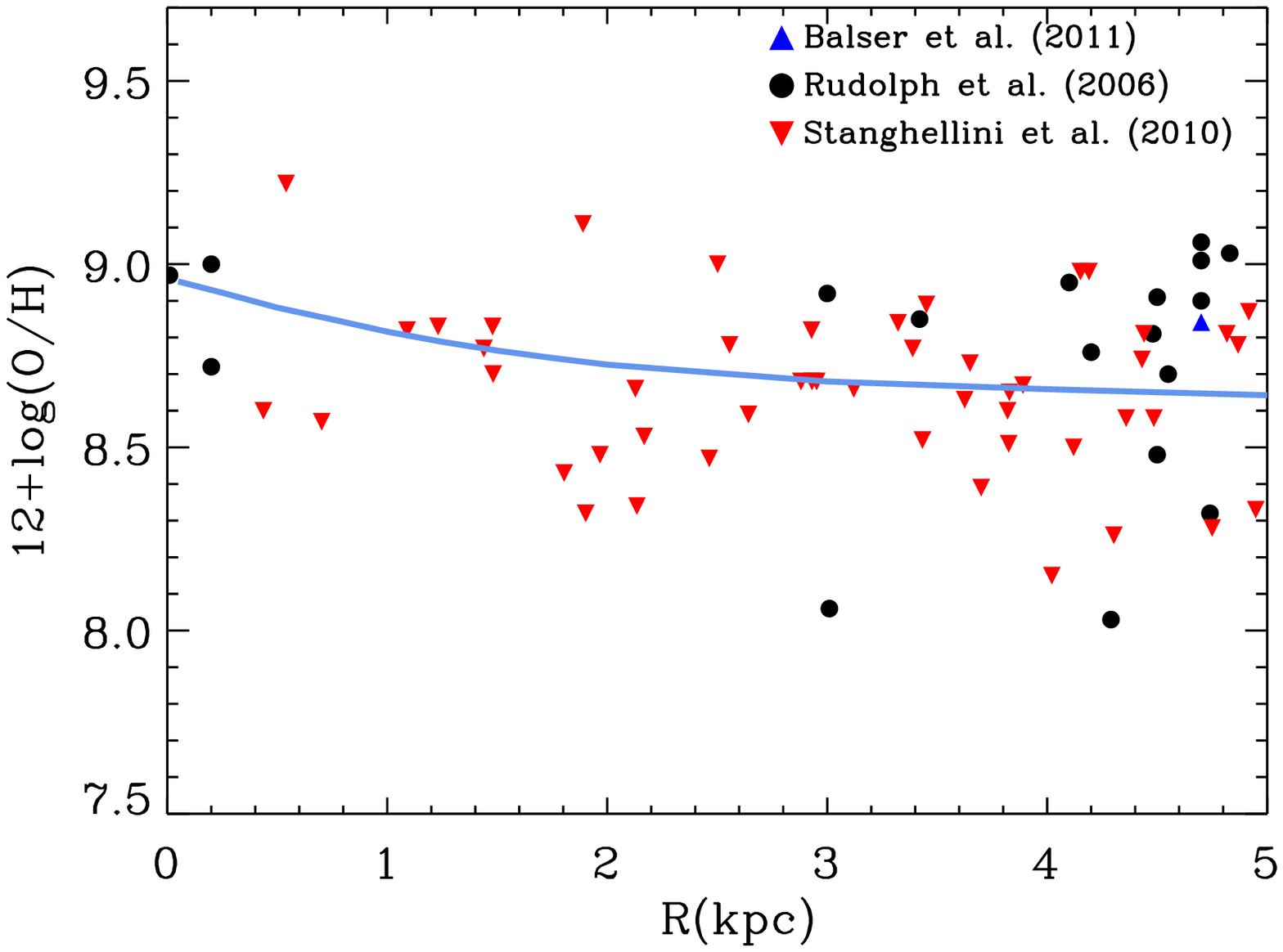}
\caption{{\it Left:} The infall of gas in the GB as a function of the redshift for each $k$-th radial region, as labelled in the figure. {\it Right:} The radial O/H gradient in the GB and inner disc of the Galaxy. Data are from HII regions and planetary nebulae, as labelled.}
\label{fig1}
\end{center}
\end{figure}

\section{Results}

\label{results}

The results of the simulations for each GB $k$-th region are presented in Fig. 1 to 3.  In Fig. 1 the infall of gas is more constant in the central regions than in the outer ones. For the lower redshifts, the infall is higher in the central parts of the GB. This results in the negative radial O/H gradient observed in the right panel of the same figure.
\begin{figure}[!ht]  
\begin{center}
\includegraphics[height=8.0cm]{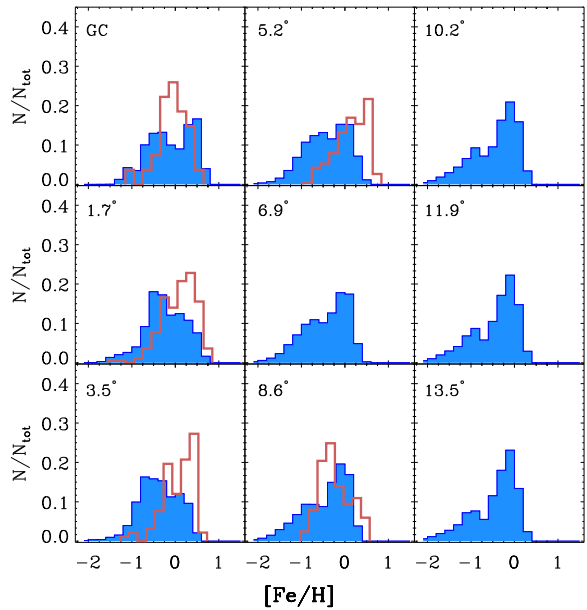}
\caption{ The GB MDF for each field from simulations (filled blue histograms) and observational data (unfilled red histograms) from Z17 and R16 (GC only).  The histograms of the simulations in the same line of sight were convolved in order to compare with the observations.}
\label{fig1}
\end{center}
\end{figure}
The GB MDF is shown in Fig. 2  and is compared with the observational data from Z17, except for the GC, where we plotted the data from \citet{ryde16}, hereafter R16. Since the observational data is a projection of stars in the line of sight, and it is given in terms of the angular distance of the GC instead of radial distance in kpc, the predictions of the model in the same line of sight were convolved to simulate the observational results. Hence, the histograms for the MDF are given for each field of the GB in degrees. The shape of the MDF depends upon the selection effects of the survey, so that we do not intend to reproduce the shapes of the MDF, but more global aspects such as the metallicity range. The model predictions for $[\alpha/{\rm Fe}]]$ ratio are displayed in Fig. 3 for different regions in the GB (left panel) and only for the central region (right panel). The models are compared with the observational data, as labelled. A higher $[\alpha/{\rm Fe}]$ ratio is predicted for the central regions, which is in agreement with the observational data (right panel). The scenario adopted for the formation of the GB may explain the spread of observational data in these figures. 

\begin{figure}  
\begin{center}
\includegraphics[height=7.5cm]{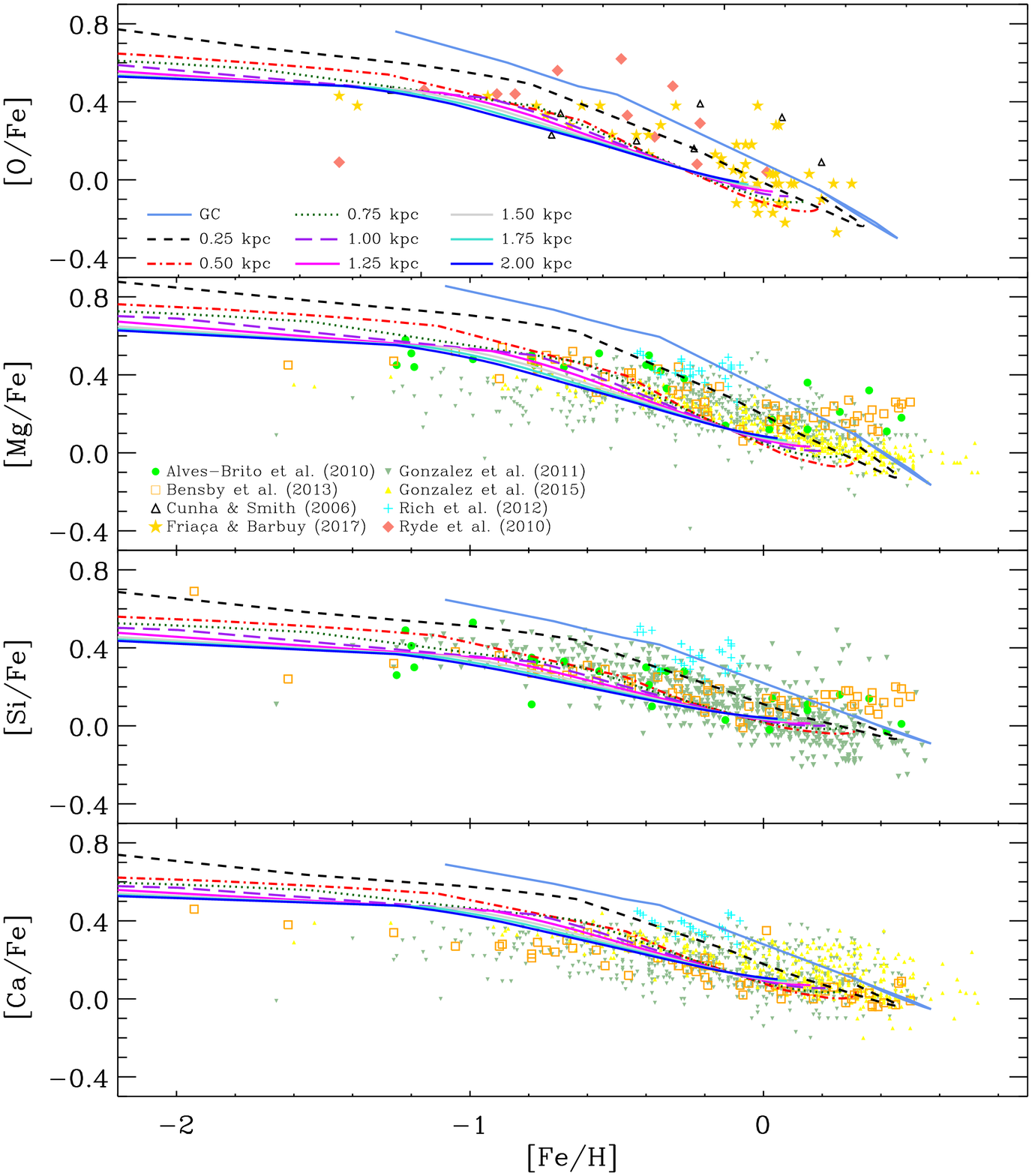}
\includegraphics[height=7.5cm]{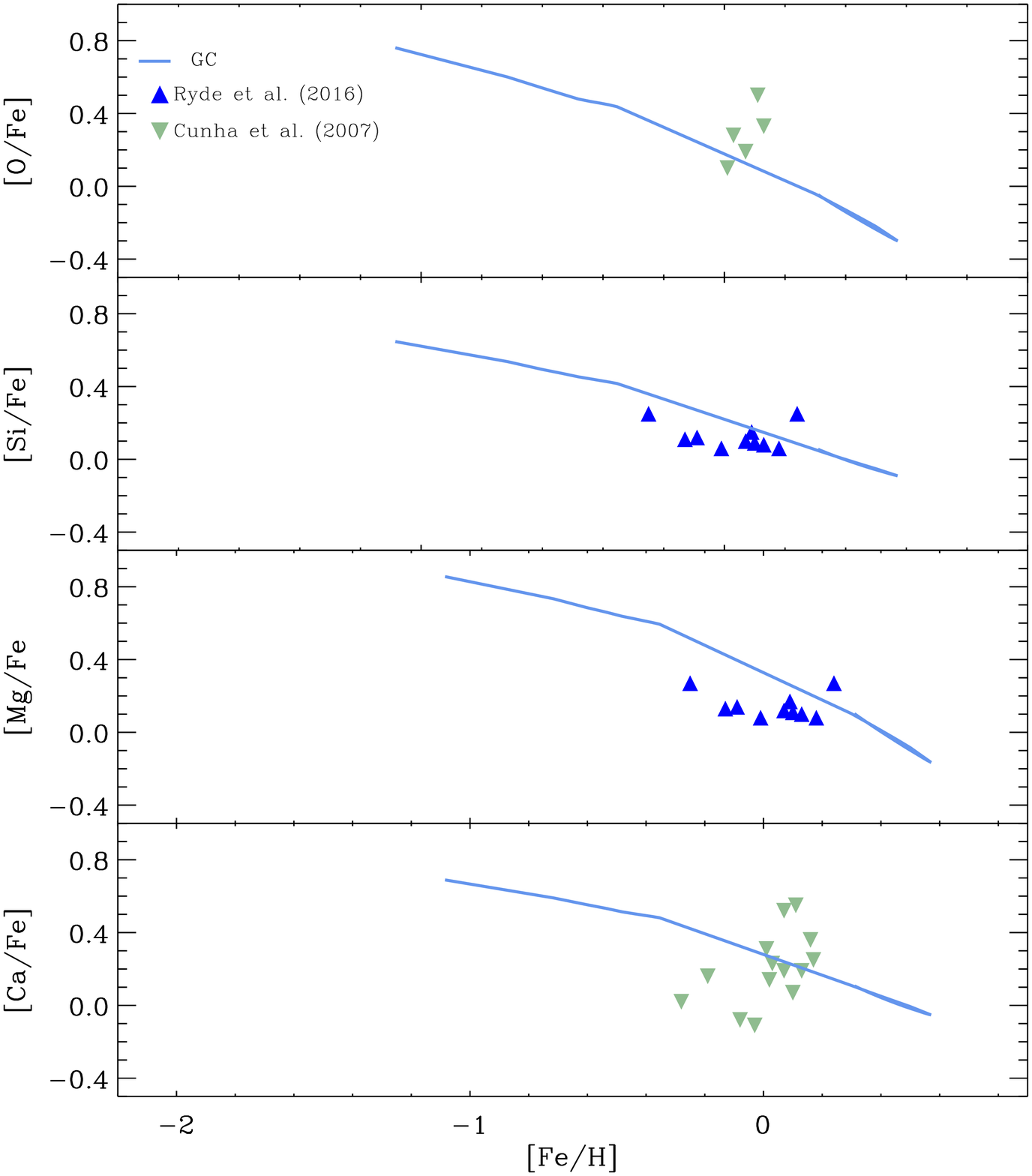}
\caption{ {\it Left:}  The [$\alpha$/{\rm Fe}] ratio from models compared with stellar observational data, as labelled. 
{\it Right:} Predictions of the simulations only for the GC contrasted with observational data inferred from the GC stars, as labelled.}
\label{fig1}
\end{center}
\end{figure}

\section{Conclusions}
\label{discussion}

The formation of the GB in an inside-out scenario, where the collapse time-scale is calculated from the GB mass distribution, is a plausible one which is able to reproduce some observational data from the literature. Most important, the simulations show a difference in the MDF depending on the GB region, as found by the observations. The models also point to a higher $[\alpha/{\rm Fe}]$ ratio in the inner regions of the GB, that reproduces the spread of stellar data.

\acknowledgments CAPES and FAPEMIG (APQ-00915-18).

\bibliographystyle{aaabib}
\bibliography{cavichia}

\newcommand{\noopsort}[1]{}
\begin{thebibliography}{}

\bibitem[\protect\astroncite{{Cavichia} et~al.}{2010}]{cavichia10}
{Cavichia} O., {Costa} R.~D.~D., {Maciel} W.~J., 2010,
\newblock {\em RMxAA}, {\bf 46}, 159

\bibitem[\protect\astroncite{{Moll{\'a}} et~al.}{2015}]{molla15}
{Moll{\'a}} M., {Cavichia} O., {Gavil{\'a}n} M., {Gibson} B.~K., 2015,
\newblock {\em \mnras}, {\bf 451}, 3693

\bibitem[\protect\astroncite{{Moll{\'a}} et~al.}{2016}]{molla16}
{Moll{\'a}} M., {D{\'{\i}}az} {\'A}.~I., {Gibson} B.~K., {Cavichia} O.,
  {L{\'o}pez-S{\'a}nchez} {\'A}.-R., 2016,
\newblock {\em \mnras}, {\bf 462}, 1329

\bibitem[\protect\astroncite{{Rojas-Arriagada} et~al.}{2014}]{rojas-arrigada14}
{Rojas-Arriagada} A., {Recio-Blanco} A., {Hill} V., {de Laverny} P.,
  {Schultheis} M., {Babusiaux} C., {Zoccali} M., {Minniti} D., {Gonzalez}
  O.~A., {Feltzing} S., {Gilmore} G., {Randich} S., {Vallenari} A., {Alfaro}
  E.~J., {Bensby} T., {Bragaglia} A., {Flaccomio} E., {Lanzafame} A.~C.,
  {Pancino} E., {Smiljanic} R., {Bergemann} M., {Costado} M.~T., {Damiani} F.,
  {Hourihane} A., {Jofr{\'e}} P., {Lardo} C., {Magrini} L., {Maiorca} E.,
  {Morbidelli} L., {Sbordone} L., {Worley} C.~C., {Zaggia} S., {Wyse} R., 2014,
\newblock {\em \aap}, {\bf 569}, A103

\bibitem[\protect\astroncite{{Sersic}}{1968}]{sersic68}
{Sersic} J.~L., 1968,
\newblock {\em {Atlas de Galaxias Australes}}

\bibitem[\protect\astroncite{{Zoccali} et~al.}{2008}]{zoccali08}
{Zoccali} M., {Hill} V., {Lecureur} A., {Barbuy} B., {Renzini} A., {Minniti}
  D., {G{\'o}mez} A., {Ortolani} S., 2008,
\newblock {\em \aap}, {\bf 486}, 177

\bibitem[\protect\astroncite{{Zoccali} et~al.}{2017}]{zoccali17}
{Zoccali} M., {Vasquez} S., {Gonzalez} O.~A., {Valenti} E., {Rojas-Arriagada}
  A., {Minniti} J., {Rejkuba} M., {Minniti} D., {McWilliam} A., {Babusiaux} C.,
  {Hill} V., {Renzini} A., 2017,
\newblock {\em \aap}, {\bf 599}, A12

\end{thebibliography}

\end{document}